\documentclass[reqno, 12pt]{amsart}

\usepackage{amsfonts}
\usepackage{amsxtra}
\usepackage{amsmath}
\usepackage{a4}
\usepackage{amssymb}
\usepackage{psfrag}
\usepackage{epsfig}
\usepackage{color}
\usepackage{epsf}
\usepackage{float}
\usepackage{a4wide}

\newtheorem{thm}{Theorem}

\newtheorem{propos}[thm]{Proposition}

\def\reff#1{(\ref{#1})}
\def\cX{{\mathcal X}} \def\cY{{\mathcal Y}} 
\def\tcX{{\tilde{\mathcal X}}}

\def\teta{{\tilde\eta}}
\def\txi{{\tilde\xi}}

\def\tA{{\tilde A}}
\def\tC{{\tilde C}}
\def\R{{\mathbb R}}  
\def\RR{{\mathbb R}}  
\def\Z{{\mathbb Z}}  
\def\ZZ{{\mathbb Z}}  
\def\cX{{\mathcal X}}

\def\square{\ifmmode\sqr\else{$\sqr$}\fi}
\def\sqr{\vcenter{
         \hrule height.1mm
         \hbox{\vrule width.1mm height2.2mm\kern2.18mm\vrule width.1mm}
         \hrule height.1mm}}                  
\def\Xnup{\cX^{n\uparrow}}

\begin{document}

\centerline{\Large Multiclass processes,}
\centerline{\Large dual points and $M/M/1$ queues}

\vskip 3mm

\centerline{\bf Pablo A. Ferrari}
\centerline{\it Universidade de S\~ao Paulo}
\vskip 2mm
\centerline{\bf James B. Martin}
\centerline{\it CNRS and Universit\'e Paris 7}


\vskip1truecm\rm
\noindent {\bf Abstract:} 
We consider the discrete Hammersley-Aldous-Diaconis process (HAD) and the
totally asymmetric simple exclusion process (TASEP) in $\Z$.
The basic coupling induces a \emph{multiclass} process which is useful in
discussing shock measures and other important properties of the processes. 
The invariant measures 
of the multiclass systems are the same for both processes,
and can be constructed as the law of the output
process of a system of multiclass queues in tandem;
the arrival and service processes of the queueing
system are a collection of independent Bernoulli product measures.
The proof of invariance involves a new coupling
between stationary versions of the processes called
a \emph{multi-line process}; this process has a collection of independent
Bernoulli product measures as an invariant measure.
Some of these results have appeared elsewhere and this
paper is partly a review, with some proofs given only in outline.
However we emphasize a new approach
via \emph{dual points}:
when the \emph{graphical construction} is used to 
construct a trajectory of the TASEP or HAD process as a function 
of a Poisson process in $\Z\times\R$, the dual points
are those which govern the time-reversal of the trajectory.
Each line of the multi-line process is governed by the dual
points of the line below.
We also mention some other processes whose multiclass
versions have the same invariant measures, and
we note an extension of Burke's theorem to multiclass queues
which follows from the results.

\paragraph{\bf MSC-class} 60K35 82C22 90B20 90B22

\def\square{\ifmmode\sqr\else{$\sqr$}\fi}
\def\sqr{\vcenter{
         \hrule height.1mm
         \hbox{\vrule width.1mm height2.2mm\kern2.18mm\vrule width.1mm}
         \hrule height.1mm}}                  

\section{Introduction}
The macroscopic evolution of the 
\emph{totally asymmetric simple exclusion process} (TASEP) \cite{Liggettbook,
  Liggettbook2} 
and the \textit{Hammersley-Aldous-Diaconis process} (HAD)
\cite{H72,AD95} can be described by the Burgers equation
\cite{Rost,Seppalainen96}.  
This equation admits shock solutions which have a
microscopic counterpart. 
A crucial tool for the study of these shocks
is Liggett's \emph{basic coupling} 
\cite{Liggettcoupling, Liggettbook}
of two or more copies of the process; see for example 
\cite{DJLStasep, Fshock, FKS, Garcia, Wick}. 
These coupled processes have a natural interpretation
as a process with two or more classes of particle
(which is also of interest from a combinatorial viewpoint
\cite{Angeltasep, DucSch}).
Such \emph{coupled} and \emph{multiclass processes},
and their invariant distributions, are the subject of this paper.

Our basic objects are the TASEP and 
a discrete-space version of the HAD process \cite{Ftagged, Guiol}.
These are continuous-time Markov processes
taking values in the state-space $\cX=\{0,1\}^\Z$;
for a configuration $\eta\in\cX$, we say 
that $\eta$ has a particle at $x$ (or that $x$ is occupied)
if $\eta(x)=1$, and that $\eta$ has a hole at $x$ 
(or that $x$ is empty) if $\eta(x)=0$.
In the TASEP, each particle tries to 
jump one site to the left at rate 1,
succeeding if the site to its left is empty.
In the HAD process,
each empty site summons at rate 1 the nearest
particle to its left.

The \emph{graphical construction} allows us to realize
either the TASEP or the HAD process
as a function of a rate 1 Poisson process on $\R\times\Z$ 
(the process of ``marks'' or ``points'' or ``bells'')
and the initial configuration.
Under the basic coupling, processes started at different
initial conditions 
$\eta^1,\eta^2,\dots,\eta^n$ are coupled by using
the same realization of the Poisson points.
Suppose the initial configurations are \emph{ordered},
in that $\eta^1(x)\leq\eta^2(x)\dots\leq\eta^n(x)$ for 
all $x\in\Z$. Then this ordering is preserved
by the dynamics of the coupling, 
and we obtain a \emph{coupled process} taking values
in the space $\Xnup$ of ordered configurations,
defined by
\begin{equation}
\label{Xnupdef}
\Xnup=\left\{
(\eta^1,\dots,\eta^n)\in\cX^n\,\,
:
\,\,\eta^1(x)\leq\dots\leq\eta^n(x)
\,\,
\text{ for all }
x\in\Z
\right\}.
\end{equation}

We define a map $R:\Xnup\mapsto\cY_n=\{1,2,\dots,n+1\}^\Z$,
taking an ordered configuration $\eta$ into a 
\emph{multiclass configuration} $\xi=R\eta$, by
\begin{equation}
  \label{Rdef}
  \xi(x) = n+1-\sum_{k=1}^n \eta^k(x).
\end{equation}
The multiclass configuration $\xi$ labels each site $i$ 
with a class. The larger
the number of $\eta$ particles at site 
$i$, the lower is its class.  
If a site is occupied in all $n$ marginals of $\eta$,
then $\xi(x)=1$ and we say that $x$ contains a first-class particle.
If $\eta^1(x)=0$ but $\eta^2(x)=1$ then $x$ contains
a second-class particle ($\xi(x)=2$), and so on.
When $x$ is empty in all the marginals of $\eta$, 
$\xi$ gives value $n+1$ at $i$. 
Conventionally, we regard such sites with value $n+1$ as holes
(this is also how they are displayed in the figures in this paper),
though they could equivalently be regarded as particles of class $n+1$.
A coupled process $(\eta_t)$ taking values in 
$\Xnup$ induces a multiclass process $(\xi_t)=(R\eta_t)$;
since $R$ is a bijection, the two are essentially equivalent.
In the case of the TASEP, the dynamics of the 
multiclass process have a natural interpretation
as a TASEP in which lower-numbered particles have
priority over higher-number particles, and can jump from the right
to displace them; these dynamics are related to various
sorting algorithms.

The Bernoulli product measures $\nu^\rho$ on $\cX$
of density $\rho\in(0,1)$
are invariant
for both the HAD process and the TASEP.
We look for invariant 
measures for the corresponding coupled processes,
whose $k$th marginal is $\nu^{\rho^k}$ for each $1,2,\dots,k$,
for fixed densities $0<\rho^1<\dots<\rho^n<1$.
The 
family of 
invariant measures
$\pi=\pi^{(n)}_{\rho^1,\dots,\rho^n}$ 
obtained 
for the coupled process is the same for
the TASEP and for the discrete HAD process, 
and also for various other particle systems with values in $\cX$.
To each such $\pi$ corresponds a distribution $\mu$ on $\cY_n$,
which is invariant for the multiclass process.
The form of $\mu$ can be obtained 
as the stationary output process of 
a system of 
multiclass $./M/1$ queues in tandem. 
A striking property of these
invariant measures is that a sample $\eta$ from $\pi$
can be obtained as a deterministic function $T$
of a configuration $\alpha\in\cX^n$ with distribution 
$\nu=\nu^{\rho^1}\times\dots\times\nu^{\rho^n}$, a product of Bernoulli
product measures.

To show that $\pi$ and $\mu$ 
are
invariant for the coupled and multiclass process respectively,
one can perform a
generator-like computation as 
\cite{Angeltasep, DJLStasep, FFKtasep, Speertasep} 
did for the 2-class TASEP. We proceed otherwise, introducing a new
process $\alpha_t=(\alpha^1_t,\dots,\alpha^n_t)$ on $\cX^n$ called
a \emph{multi-line process} and showing that 
(a) the product of product measures
$\nu$ is invariant for the multi-line process and 
(b) the process $T\alpha_t$
is the coupled process. 
See \cite{FM-tasep} for the TASEP and \cite{FM-had}
for an analogous construction 
for the continuous-space HAD process. 
We review the method in this paper,
and describe an approach to the multi-line process
via \emph{dual points}, rather different to the approach of \cite{FM-tasep}.

Stationary versions of the 
TASEP and the HAD process can be constructed as deterministic
functions of the Poisson process $\omega$. 
For each density $\rho$ and 
almost all Poisson point configurations $\omega$, 
there exists a unique trajectory of the 
process $(\eta_t,\,t\in\R)$ with time-marginal
distribution $\nu^\rho$ and governed by $\omega$. This was proved for the
continuous-space HAD process 
in \cite{FM-had} and the proof extends to the cases discussed
here. 
The $\omega$ points and the trajectory of the process induce new points
$D_\rho(\omega)$, defined --- roughly speaking --- as the points governing the
time-reversed trajectory. They are called \emph{dual points} and,
just as $\omega$ itself, form a Poisson process on $\R\times\Z$.
In addition, the dual points with time coordinate less than $t$ are independent
of the particle configuration at time $t$. 
These properties have been shown for the
continuous-space version of the HAD by Cator and Groeneboom \cite{CG}; see also
\cite{FM-had}.  We show them here for both the TASEP and the discrete HAD
process.  The proof is a bit more complicated because a trajectory 
in the discrete-space case 
does not determine all the Poisson points governing it. To overcome 
this problem we
augment the state-space and introduce 
spin-flip processes to mark the Poisson
points missed by the trajectories. See Proposition \ref{dp} for the HAD and
Proposition
\ref{c11} for the TASEP.  These are the main new results of this paper.

One way to construct a stationary trajectory $(\alpha_t,t\in\RR)$ 
of the multi-line process referred to above,
whose marginal distribution at any fixed time $t$ 
is $\nu=\nu^{\rho^1}\times\dots\times\nu^{\rho^n}$,
is as follows.  
The bottom line $(\alpha^n_t, t\in\RR)$ of the 
process is constructed as a the unique trajectory with
marginal $\nu^{\rho^n}$ governed by the Poisson points $\omega$.  
The dual
points $\omega^{n-1}$ of the bottom trajectory and the density $\rho^{n-1}$
are then used to construct the $(n-1)st$ line, 
and so on. Since $\alpha^k_t$, the $k$th marginal at
time $t$, depends 
only on the Poisson points $\omega^k$ with time coordinate less
than $t$, which are independent of $\alpha^{k-1}_t,\dots,\alpha^1_t$,
the resulting distribution of $\alpha_t\in\cX^n$ is indeed
the product distribution $\nu$. The process 
is time-invariant by construction.

In the case of the 
continuous-space HAD process,
the multi-line process is closely related to the 
\emph{polynuclear growth model} studied by Pr\"ahofer and Spohn
\cite{PS00,PS02}; see also \cite{Patrik04,fp05}. 
In those papers the points in
$\omega$ and the dual points are called \emph{nucleation events} and
\emph{annihilating events} respectively; see in particular Figure 2 in
\cite{Patrik04}.

In Section \ref{sim} we construct the 
coupled invariant measure $\pi$ and the
multiclass invariant 
measure $\mu$ as functions of a product of Bernoulli product measures. 
In Section \ref{shad} we review the proof of the invariance of 
$\pi$ (or $\mu$)
for the coupled (respectively, multiclass) HAD process; 
at the end of this section
we mention a case of the \emph{long range exclusion process} 
which is equivalent to the HAD process.  
In Section \ref{stasep} we review the proof for the case of the TASEP.
In Section \ref{sdynamics} we give some examples of other
dynamics for which the associated multiclass processes
have the family of measures $\mu$ as invariant distribution,
including certain ``sequential'' TASEPs defined in discrete time.
We also mention various related cases for which
$\mu$ is \emph{not} invariant.
Finally, in Section \ref{sburke}, we note a 
multiclass generalization of 
\emph{Burke's theorem}. The measures $\mu$
constitute \emph{fixed point arrival processes} for a 
multiclass priority $./M/1$ queue, in the sense that
the law of the output process of the queue is the same as 
the law of the input.
This property can be deduced from the invariance of $\mu$
and the tandem queue construction used in the proof
(although more direct proofs are also available; see \cite{MarPra}).

In summary, our main aims in this paper are as follows:
(i) to review the results of \cite{FM-tasep} 
and \cite{FM-had} describing invariant measures
of multiclass processes;
(ii) to illustrate the application of the results and methods 
to various different particle systems;
(iii) to describe an approach via \emph{dual points}
which adapts well to various different processes and 
which differs, for example, from the more specific
arguments used in \cite{FM-tasep} for the TASEP case,
and (iv) to emphasize the correspondence between
the multiclass process with values in $\cY_n$ 
and the coupled process
with values in $\Xnup$.

\section{Multiclass invariant measures}
\label{sim}

In this section we will construct a family 
of measures on $\Xnup$, which we will later
show to be invariant for the coupled HAD and TASEP processes.
For given particle densities $0<\rho^1<\dots<\rho^n<1$,
we will construct a measure
$\pi=\pi^{(n)}_{\rho^1,\dots,\rho^n}$
on $\Xnup$ whose $k$th marginal will have distribution 
$\nu^{\rho^k}$,
Bernoulli product measure of density $\rho^k$
(that is, the measure under which 
each site is occupied independently
with probability $\rho^k$).

Let $\eta$ have distribution $\pi$,
and write $\xi=R\eta\in\cY_n$ for
the configuration obtained from the map defined at \reff{Rdef}.
Since $\pi$ is invariant for the coupled processes,
the distribution $\mu=R\pi$ of $\xi$ will be invariant 
for the corresponding multiclass processes.

The invariant measure $\pi$ on $\Xnup$
is constructed
starting from product measure 
$\nu=\nu^{\rho^1}\times\dots\times\nu^{\rho^n}$
on $\cX^n$. Let $\alpha=(\alpha^1,\dots,\alpha^n)\in\cX^n$ 
be distributed according to this product measure $\nu$. 
We will interpret the particles of $\alpha$ as events
in a system of queues in tandem in discrete time; the sites of $\ZZ$ 
now correspond to times in the queueing system.

\subsection{Construction of the coupled invariant measure $\pi$}
\label{two}

Consider a queueing server in discrete time,
governed by $\alpha^1$ and $\alpha^2$ in the following way.
The particles of $\alpha^1$ represent 
times at which a customer arrives at the queue,
and the particles of $\alpha^2$ represent
potential service times 
(that is, times at which a customer can depart from the
system, if any are present).
At time $i$, the queue length increases
by 1 if $\alpha^1(i)=1$ and $\alpha^2(i)=0$;
it stays the same if $\alpha^1(i)=\alpha^2(i)$;
and it decreases by 1 if $\alpha^1(i)=0$ and 
$\alpha^2(i)=1$, unless it was already 
0 in which case it remains 0.
Then if $Z(j)$ is the queue length just after time $j$,
one has
\begin{equation}
  \label{a20}
  Z(j)  = (Z(j-1) + \alpha^1(j) - \alpha^2(j))^+ \,,\quad j\in\Z.
\end{equation}
Since $\alpha^1$ and $\alpha^2$ 
are distributed as independent Bernoulli product measures,
with densities $\rho^1$ and $\rho^2$ respectively,
and $\rho^1<\rho^2$, 
the queue-length process $Z$ is a 
positive recurrent Markov chain 
(in fact, its stationary distribution is 
geometric with parameter $\rho^1/\rho^2$),
and there is an essentially unique way 
to construct a stationary process $Z$ as a function
of the input $\alpha$, namely by
\begin{equation}
\label{Zdef}
Z(j)=\sup_{r\leq j}
\left(
\sum_{i=r}^j
\left[\alpha^1(i)-\alpha^2(i)\right]
\right)_+ .
\end{equation}
The evolution of $Z(j)$ is illustrated 
in Figure \ref{f5}. 
\begin{figure}[htb]
\begin{center}
\input{loynes.pstex_t}  
\caption{Construction of $(Z(j))$. 
The queue is empty at the moment of the first
  service time. In the plot of $Z$ unused services are indicated with a small
  vertical mark and arrivals served instantaneously with a long vertical
  segment.}
\label{f5}
\end{center}
\end{figure}
It can be seen that if an arrival and a potential service occur at the same
time, a customer may spend no time in the system; this is the case of the
last customer to arrive.
 coupled configuration is 

The configuration $D=D(\alpha^1,\alpha^2)$, representing the 
departure times from the queue, is then defined by
\begin{equation}
\label{Ddef}
D(j)=
\begin{cases}
1&\text{if } \alpha^2(j)=1 \text{ and } Z(j-1)+\alpha^1(j)>0
\\
0&\text{otherwise.}
\end{cases}
\end{equation}

The interpretation is that $D(j)=1$ if a customer departs from 
the queue at time $j$.
Allowing the value $\infty$ in \reff{Zdef},
this definition of the operator $D:\cX^2\mapsto\cX$ 
makes sense for any $\alpha^1$ and $\alpha^2$,
but in fact we will only use it in 
cases where $\alpha^1$ and $\alpha^2$ 
have independent Bernoulli product measures
of densities $\rho^1$ and $\rho^2$ with $\rho^1<\rho^2$.
Then the queueing process is stable
(since the rate of arrivals is lower than the rate of services).
In queueing theory terminology the queue is called a 
\emph{discrete-time $M/M/1$ queue} (where 
``1'' indicates a single-server queue and ``$M$'' stands for \emph{memoryless},
indicating that the arrival and service processes each
have product measure).

We note the following useful properties of the operator $D$.
The first two follow immediately from \reff{Zdef} and \reff{Ddef},
while the third is Burke's Theorem for a discrete-time $M/M/1$ queue
\cite{HsuBurke}:
\begin{propos}
\label{Dthings}
$ $

\begin{itemize}
\item[(i)] $D(\alpha^1,\alpha^2)\leq \alpha^2$.
\item[(ii)] If $\tilde{\alpha}^1\leq \alpha^1$
then $D(\tilde{\alpha}^1,\alpha^2)\leq D(\alpha^1,\alpha^2)$.
\item[(iii)] 
If $\alpha^1$ and $\alpha^2$ 
have independent Bernoulli product measures
of densities $\rho^1$ and $\rho^2$ with $\rho^1<\rho^2$,
then $D(\alpha^1,\alpha^2)$ also has 
Bernoulli product measure with density $\rho^1$.
\end{itemize}
\end{propos}

We now define a sequence of operators
$D^{(n)}:\cX^n\mapsto\cX$ as follows.
Let $D^{(1)}(\alpha^1)=\alpha^1$,
and 
then recursively for $n\geq2$, let
\begin{equation}
\label{Dndef}
D^{(n)}(\alpha^1,\alpha^2,\dots,\alpha^n)=
D\left(
D^{(n-1)}(\alpha^1,\dots,\alpha^{n-1}),\alpha^n
\right).
\end{equation}
The configuration $D^{(n)}(\alpha^1,\alpha^2,\dots,\alpha^n)$
represents the departure process from a system of 
$(n-1)$ queues in tandem. The arrival process to the first
queue is $\alpha^1$. The service process of the $k$th 
queue is $\alpha^{k+1}$, for $k=1,\dots,n-1$.
Finally, for $k=2,\dots,n-1$, 
the arrival process to the $k$th queue
is given by the departure process of the $(k-1)$st queue.
This is known as a system of $./M/1$ queues in tandem.

Note $D^{(2)}(\alpha^1,\alpha^2)=D(\alpha^1,\alpha^2)$.
By applying Proposition \ref{Dthings}
repeatedly, we obtain the following properties of $D^{(n)}$:
\begin{propos}
\label{Dnthings}
$ $

\begin{itemize}
\item[(i)] 
$D^{(n)}(\alpha^1,\dots,\alpha^n)
\leq 
D^{(n-1)}(\alpha^2,\dots,\alpha^n)
\leq
\dots
\leq
\alpha^n$.
\item[(ii)] 
If $\alpha^1,\dots,\alpha^n$ 
have independent Bernoulli product measures
of densities 
\\
$\rho^1<\dots<\rho^n$,
then $D^{(n)}(\alpha^1,\dots,\alpha^n)$ also has 
Bernoulli product measure with density $\rho^1$.
\end{itemize}
\end{propos}
Now define the configuration $\eta=(\eta^1,\dots,\eta^n)$ by
\begin{equation}
\label{Tdef}
\eta^k=D^{(n-k+1)}(\alpha^k,\alpha^{k+1},\dots,\alpha^n).
\end{equation}
From Proposition \ref{Dnthings} we have that
\begin{itemize}
\item[(i)]$\eta\in\Xnup$ (that is, $\eta^k\leq\eta^{k+1}$
for all $k=1,\dots,n-1$);
\item[(ii)] for each $k$, $\eta^k$ has marginal distribution $\nu^{\rho^k}$.
\end{itemize}
We then define the map $T:\cX^n\mapsto\Xnup$
by $T\alpha=\eta$. The desired distribution $\pi$ on $\Xnup$ is the 
induced distribution of $\eta$ (that is, $\pi=T\nu$).

\subsection{Construction of the multiclass measure $\mu$}

Let $\eta$ have the distribution $\pi$ constructed above,
which we will show to be invariant for the coupled HAD and TASEP
processes. Let $\xi=R\eta$, where $R$ is the map defined at \reff{Rdef}.
Then $\mu=R\pi$, the distribution of $\xi$,
will be invariant for the multiclass HAD and TASEP processes.

This multiclass invariant measure can also be described directly
via a tandem queueing system with multiclass queues. 
(This direct construction described below is not 
necessary to understand the proofs of invariance
in later sections, which are written in terms of the construction
of $\pi$ given in the previous section).

As before, the system will now contain $n-1$ queues. 
The arrivals to the first queue are again the 
particles of $\alpha^1$. 
For $k=2,\dots,n-1$,
the arrivals to the $k$th queue
correspond to the services of the $(k-1)$st queue,
and are given by the particles of $\alpha^k$.
These are partitioned into $k-1$ different classes; these classes 
will be served by the $k$th queue 
according to a \emph{priority policy},
under which lower-numbered classes are served
ahead of higher-numbered classes.
A class $r$ customer departing from queue $k-1$
becomes a class $r$ customer arriving at queue $k$;
an unused service at queue $k-1$ becomes a class $k$
customer arriving at queue $k$.
Finally, the services of the $(n-1)$st queue
are given by the particles of $\alpha^n$, and will
be assigned $n$ different classes; 
this will yield the $n$-type multiclass configuration desired,
distributed according to $\mu$.

The partition of the particles of $\alpha^k$ into $k$ classes
is written using configurations $\beta_1^k,\dots,\beta_k^k$ 
such that $\beta_1^k+\dots+\beta_k^k=\alpha^k$.
The configuration $\beta_r^k$ represents
the $\alpha^k$-particles of class $r$, for $r=1,2,\dots k$.

Of course $\beta^1_1=\alpha^1$.
Then for $2\leq k\leq n$, 
we set 
\begin{align*}
\beta_1^k&=D(\beta_1^{k-1}, \alpha^k);
\\
\beta_r^k&=D(\beta_r^{k-1}, \alpha^k-\beta_1^k-\dots-\beta_{r-1}^k)
\,\,\text{ for } 1<r<k;
\\
\beta_k^k&=\alpha^k-D(\alpha^{k-1}, \alpha^{k}).
\end{align*}

When a customer departs from queue $k$ 
(that is, when an $\alpha^k$-service occurs
and there is at least one customer present),
the customer which departs is the lowest-numbered
one present (including one which may just have arrived).
From the equations above for the $\beta_r^k$, 
this may be understood as follows:
the $r$th-class customers arriving 
at queue $k$ experience a service process
which corresponds to $\alpha^k$ but with 
the service times used by customers of classes $1,\dots,r-1$ removed.

Now the multiclass configuration $\xi$ is
derived from the $n$th line in the natural way;
for $k=1,2,\dots,n$,
let $\xi(j)=k$ if $\beta_k^n(j)=1$;
otherwise (i.e.\ if $\alpha^n(j)=0$)
let $\xi(j)=n+1$.
Call \[\xi=M\alpha \in\cY_n=\{1,\dots,n+1\}^\Z\] the
resulting multiclass configuration.

\begin{figure}[htb]
\begin{center}
\input{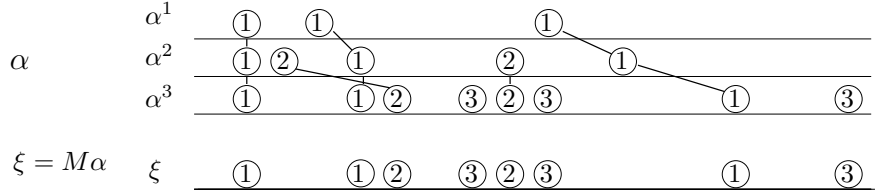}  
\caption{Construction of 
$\xi=M\alpha$. Here $n=3$. 
Each particle on line $k$ represents 
a site $i$ such that $\alpha^k(i)=1$; 
the particle carries the label $r$ such that $\beta_r^k(i)=1$
(representing a particle of $r$th class). In this diagram and
later, sites in the multiclass configuration $\xi$
with value $n+1$ are treated as holes and left empty.}
\label{multiclass}
\end{center}
\end{figure}

The construction is most easily understood from a picture;
see Figure \ref{multiclass} for an example with $n=3$.

\begin{figure}[htb]
\begin{center}
\input{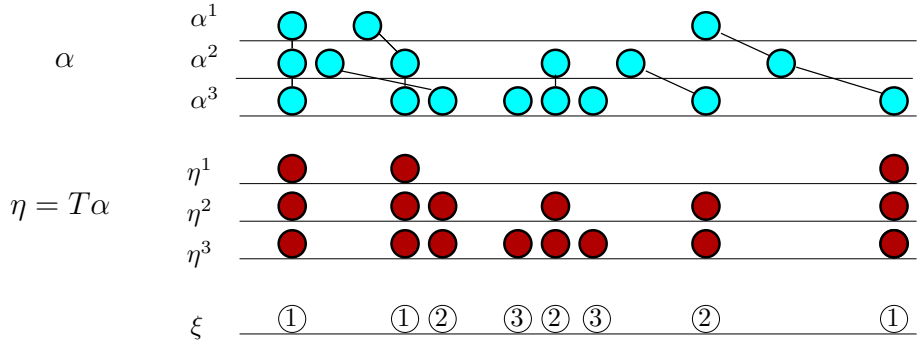} 
\caption{From independent to ordered configurations}
\label{order} 
\end{center}
\end{figure}
The relations between the multiclass configuration $\xi$, the multi-line
configuration $\alpha$ and the ordered (or coupled) configuration 
$\eta=T\alpha$ is given by 
\begin{equation}
  \label{b2}
  \xi=R\eta=R(T\alpha)=M\alpha.
\end{equation}
This correspondence is illustrated in Figure \ref{order}.
To establish 
formally this equivalence between the construction of the function
$M$ here and the construction of the function $T$ in the previous section,
one can check by induction that for any $1\leq r\leq k\leq n$,
\[
\beta_1^k+\beta_2^k+\dots+\beta_r^k=
D^{(k-r+1)}(\alpha^r,\dots,\alpha^k).
\]

Define $\mu=M\nu=RT\nu$. This is the distribution of $\xi=M\alpha$ when
$\alpha$  
consists of $n$ independent
configurations with Bernoulli product distributions of parameters
$0<\rho^1<\dots<\rho^n<1$.

\subsection{The case $n=2$}

We give a few words about the constructions above in the 
case $n=2$. We now have a single queue with arrival process
$\alpha^1$ and service process $\alpha^2$
(that is, a single $M/M/1$ queue).
The particles of $\eta^2$ are simply the particles of $\alpha^2$,
that is, the potential service times.
The particles of $\eta^1$ are the subset of those
times where a departure actually occurs, 
namely the particles of $D(\alpha^1,\alpha^2)$.
The \emph{discrepancies} between the two 
configurations $\eta^1$ and $\eta^2$
correspond to second-class particles in the two-class interpretation;
that is, to sites $j$ where $\xi(j)=2$.
These correspond to \emph{unused services} in the queue.

This construction was described by Angel \cite{Angeltasep} 
for the invariant measure of the two-class
TASEP. The queueing interpretation can be found in Ferrari and Martin
\cite{FM-tasep}. 
The measure so obtained has been first computed by Derrida, Janowsky, Lebowitz
and Speer \cite{DJLStasep} and then described in other
ways by \cite{FFKtasep, Speertasep, DucSch} 
for the two-class TASEP.

\section {Discrete HAD}
 \label{shad}
 The discrete-space 
Hammersley-Aldous-Diaconis process is a continuous-time 
Markov process taking values in a subset of $\cX$.
At rate one each site $j$ calls the closest particle to the left of $j$
(including $j$) making it jump to $j$. The generator of the process is
\begin{equation}
  \label{a1}
  L_Hf(\eta) = \sum_{j\in\Z} [f(A_j\eta) - f(\eta)]
\end{equation}
where, writing $i= i(\eta,j)=\max\{k\le j:\, \eta(k) = 1\}$ 
for the closest occupied site of $\eta$ to the left of $j$,
\begin{equation}
  \label{a2}
  A_j\eta(k) = 
\left\{
\begin{array}{ll} 
\eta(k) &\hbox{ if } k\neq i,j 
\\ 
1&\hbox{ if } k=j\\
0&\hbox{ if } k=i \text{ and } i<j
\end{array}\right. 
\end{equation}
\begin{figure}[htb]
\begin{center}
\input{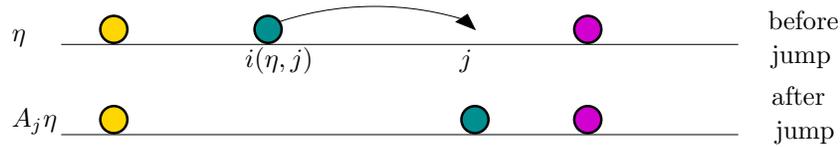}  
\caption{Discrete Hammersley process. Jumps occur at rate 1} 
\label{f1}
\end{center}
\end{figure}

\begin{figure}[htb]
\begin{center}
\input{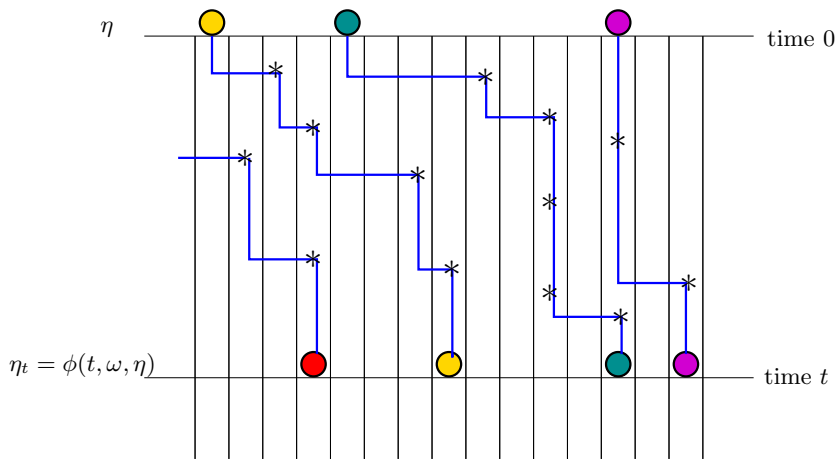}  
\caption{Harris construction} 
\label{f2}
\end{center}
\end{figure}

\paragraph{\bf Harris graphical construction}
To construct the process we attach to each site an independent 
Poisson process of rate 1;
these processes form a rate-1 Poisson process on $\R\times\Z$. 
Bells ring at the
points (or space-time events)
of this process, which 
are represented by * in Figure \ref{f2}. The space of point configurations is
called $\Omega$ and single 
point configurations are called $\omega$. When a bell
rings at $j$ (that is, at a time $s$ such that $(j,s)\in\omega$), 
the closest particle to the left jumps to $j$. A possible point
configuration and the resulting trajectory are 
illustrated in Figure \ref{f2}.
Such a construction is easily seen to be well-defined
for the corresponding process in a finite region
(since there are finitely many points of $\omega$
in any finite time-interval). 
To define the process on all of $\ZZ$,
we need to exclude the possibility of particles
escaping immediately to $+\infty$, and should restrict
to the state space 
\[
\tcX=\left\{
\eta\in\cX:
\lim_{r\to\infty}
r^{-1/2}\sum_{j=1}^r\eta(j)=\infty
\right\}.
\]
For details, see Sepp\"al\"ainen \cite{Seppalainen96},
where the analogous construction for the continuous-space HAD
is carried out.

If one calls $\omega$ a realization of the points, then the configuration at
time $t$ of the process is a function called $\phi$ of $\omega$ and the initial
configuration $\eta_0$: 
\begin{equation}
  \label{a4}
  (t,\omega,\eta_0)\mapsto\phi(t,\omega,\eta_0)
\end{equation}.

For given $\eta_0$ and $\omega$, the process
$(\eta_t,t\geq0)$ defined by $\eta_t=\phi(t,\omega,\eta_0)$
is called the HAD process 
\textit{governed by $\omega$ with initial condition $\eta_0$}.
In fact one has
\begin{equation}
\label{st}
\eta_t=\phi(t-s,\omega,\eta_s)
\end{equation}
for all $0\leq s<t<\infty$. 

The invariant measures of the HAD process are the
Bernoulli product measures $\nu^\rho$ with density $\rho\in(0,1]$ 
(and mixtures of them).

Using the Kolmogorov extension theorem, we can dispense with the initial
condition and construct jointly the Poisson points $\omega$ and an evolution
$(\eta_t, t\in \RR)$ such that, for all $t$, the marginal distribution of
$\eta_t$ is $\nu^\rho$, and which satisfies \reff{st} for all
$-\infty<s<t<\infty$.  We again say that $(\eta_t,t\in\RR)$ is governed by
$\omega$.  In fact, it turns out that the construction of such a bi-infinite
trajectory is essentially unique, as soon as the particle density $\rho$ is
fixed:

\begin{propos}
\label{uniqpropos}
Let $\rho\in(0,1)$.
Then there exists an essentially unique function $H_\rho$ 
mapping elements $\omega$ of $\Omega$ to trajectories
$(\eta_t,t\in\R)$ such that:
\begin{itemize}
\item[(i)]
The induced law of $(\eta_t,t\in\R)=H_\rho(\omega)$ is stationary in time.
\item[(ii)]
The marginal law of $\eta_t$ for each $t$ is space-ergodic with particle density
$\rho$.
\item[(iii)]
With probability 1, $(\eta_t,t\in\R)$ is a HAD evolution governed by $\omega$.
\end{itemize}
(Here ``essentially unique'' means that if $H'_\rho$ is another 
function satisfying the three conditions, then $H_\rho(\omega)=H'_\rho(\omega)$
with probability 1).
Then in fact the marginal law of $\eta_t$ for each $t$ is $\nu^\rho$.
\end{propos}

Proposition \ref{uniqpropos} can be proved following the
approach of Ekhaus and Gray \cite{eg94}; 
see Mountford and Prabhakar \cite{mp95}
and our proof for the continuous-space HAD in \cite{FM-had}.
One might conjecture that a stronger statement holds:
for almost all $\omega$, there exists a unique HAD trajectory $(\eta_t,t\in\R)$ 
governed by $\omega$ such that, for all $t$,  the configuration $\eta_t$ 
has particle density $\rho$.

\paragraph{\bf Coupling}
Different initial configurations $\eta^1_0,\dots,\eta^n_0$ 
with the same Poisson
bells $\omega$ 
produce a joint process whose marginals are the HAD process
with those initial configurations:
 \[\eta^k_t = \phi(t,\omega,\eta^k_0 )\]

Hence for an initial condition $\eta_0=(\eta^1_0,\dots,\eta^n_0)\in\tcX^n$,
we can describe the coupled HAD process by
\[
\eta_t=\phi^{(n)}(t,\omega,\eta_0),
\]
where the function $\phi^{(n)}:\RR\times\Omega\times\tcX^n\mapsto\tcX^n$
is defined by
\[
(\phi^{(n)}(t,\omega,\eta))^k=\phi(t,\omega,\eta^k).
\]
The generator of the coupled process is given by
\begin{equation}
  \label{b6}
  L_Cf(\eta) = \sum_{j\in\Z} [f(C_j\eta) - f(\eta)]
\end{equation}
where
\begin{equation}
  \label{p98}
 C_j\eta= (A_j\eta^1,\dots,A_j\eta^n).
\end{equation}
for $A_j$ defined in \reff{a2}.

\begin{figure}[htb]
\begin{center}
\input{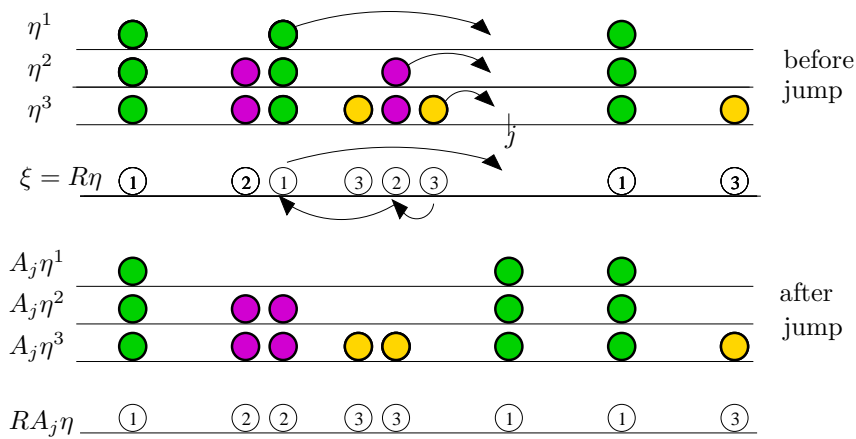}  
\caption{Coupled and multiclass processes. Effect of a bell at $j$}
\label{f4}
\end{center}
\end{figure}

If the initial configurations are ordered, that is,  $\eta^k_0(i)\le
\eta^{k +1}_0(i)$, for all $i$, then
$\eta^k_t\le\eta^{k+1}_t$ for all $t$. 
Put another way, we can regard the coupled process
as a process taking values in the space $\Xnup$ 
of ordered configurations.
In Figure \ref{f4} we illustrate
the jumps produced by a bell at site $j$ in such a case. 
The closest particle to the left of
$j$ in each marginal jumps to $j$; the jumps are simultaneous.

\paragraph{\bf Multiclass process} 
From now on we indeed regard the coupled process 
as a process taking values in the space $\Xnup$ of ordered
configurations.
Now we can regard as first-class
particles the sites occupied in all marginals, second-class particles those
occupied from the second marginal but not the first, and so on.
The \emph{multiclass process} is defined in terms of the coupled 
process by $\xi_t:=R\eta_t$. 
Since $R$ is a bijection from $\Xnup$ to $\cY_n$.
this is also a Markov
process (whose behavior is not completely intuitive;
compare for example with the more natural behavior of the multiclass
TASEP process considered in Section \ref{stasep}).
The generator of the multiclass process
can be written in terms of the generator of the coupled process
by $L_{MC}=RL_CR^{-1}$, and the operator
$R$ commutes with the dynamics of the coupled and multiclass processes.
Figure \ref{f4} illustrates the correspondence between the two processes.

\subsection{Invariance of $\mu$}

\begin{thm}
\label{t1}
Let $\alpha=(\alpha^1,\dots,\alpha^n)$ have law $\nu$, product of Bernoulli
product measures with densities $\rho^1<\dots<\rho^n$.  Then 
$\pi$, the law of
$T\alpha$, is invariant for the coupled HAD process $(\eta_t)$ and $\mu$,
the law of $M\alpha$, is invariant for the multiclass HAD process~$(\xi_t)$.
\end{thm}

\paragraph{\bf Sketch of proof} 
From \reff{b2}, the statements are equivalent. Hence it
suffices to show that the law of $\eta=T\alpha$ is invariant for the coupled
process.  We do it in two steps. First introduce new dynamics $\alpha_t =
(\alpha^1_t,\dots,\alpha^n_t)$ called 
the \emph{multi-line process}, and then show:

1) The product measure $\nu$ is invariant for the multi-line process
$\alpha_t$.

2) $T\alpha_t$ is the coupled process $\eta_t$.

\noindent These statements are Propositions \ref{w2} and \ref{w1} below.  
\square

In fact one can go on to show that
this family of measures $\mu$,
indexed by the densities $\rho^1,\dots,\rho^n$,
are the \emph{only} extremal invariant measures
for the multiclass process. The proof
of such a result follows a coupling argument
of Ekhaus and Gray \cite{eg94} as implemented by
Mountford and Prabakhar \cite{mp95}. See 
\cite{FM-had} for the argument for the continuous-space HAD process.

\subsection{Dual points}

Given a particle density $\rho\in(0,1)$
and a realisation of the Poisson marks $\omega$,
Proposition \ref{uniqpropos} provides
a stationary HAD trajectory $(\eta_t)=H_\rho(\omega)$ 
governed by $\omega$ with time-marginal $\nu^\rho$.

We define another set of marks 
$\Delta_\rho(\omega)$, called \emph{dual points}.
These are given by the positions of the particles 
just before jumps:
\begin{equation}
  \label{dp1}
\Delta_\rho(\omega) := \{(i(\eta_{t-},j),t):\, (j,t)\in\omega\}
\end{equation}
where $i(\eta,j)$ is the position of the closest $\eta$ particle to the left of
$j$, as defined after \reff{a1}. This includes ``jumps'' of null size, when
$i=j$.

\begin{figure}[htb]
\begin{center}
\input{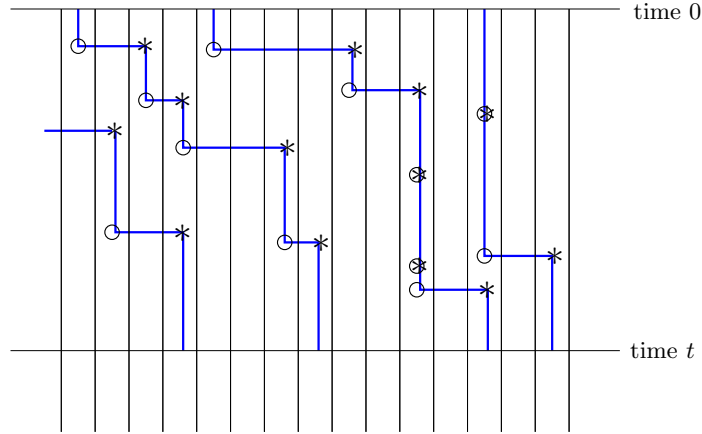}  
\caption{The dual points of the trajectory of Figure \ref{f2} 
are represented by circles.}
\label{ml1b}
\end{center}
\end{figure}

The points $\omega$ and the dual points $\Delta_\rho(\omega)$ 
are illustrated in Figure \ref{ml1b}.
Since the dual points are located in the space-time
positions just vacated by particles,
they govern the time-reversal of the trajectory.
More precisely: the time-reversed and space-reversed
trajectory is a HAD trajectory governed by
(the time- and space-reversal of) $\Delta_\rho(\omega)$.
In visual terms: turning Figure \ref{ml1b} upside-down
exchanges the roles of the stars and the circles.

The law of the dual points $\Delta_\rho(\omega)$ is then also
Poisson, just as the law of $\omega$ itself. 
This is the first part of the following result.
\begin{propos}
  \label{dp}
Let $\omega$ be a Poisson process in 
$\R\times\Z$, and let $\Delta_\rho(\omega)$ be the dual 
points for the HAD trajectory with particle density $\rho$
governed by $\omega$.
Then $\Delta_\rho(\omega)$ is also a Poisson process in $\R\times\Z$.
Furthermore
$\{(x,s)\in \Delta_\rho(\omega):\, s<t\}$, 
the set of dual points earlier than 
$t$, is independent of the configuration $\eta_t$.
\end{propos}

\paragraph{\bf Proof} The proof is in the spirit of Reich's \cite{Reich} proof
of Burke's theorem, used by Cator and Groeneboom \cite{CG} for the continuous
space HAD. 
The idea is to consider the time-reversal of the process, 
and goes as follows.
As commented above, the dual points govern the reverse process.
By doing a generator calculation
(or verifying an equivalent detailed-balance property)
one obtains that the time-reversal
of the equilibrium HAD process with density $\rho$
is again an equilibrium HAD process with density $\rho$, but now
with jumps to the left.
(Put another way, the time-reversal 
of the process has the same law as the space-reversal).
Now we would like to conclude that that the dual points 
therefore must also be Poisson.
However, in the discrete-space case, the 
problem is that 
the trajectory of the HAD process does
not identify all the points which govern it;
it is also necessary to keep track of the 
points producing null jumps, which are not visible from the
trajectory alone.
To overcome this, we will add an auxiliary spin-flip process.

Let $\gamma_t\in\cX$ be the process which behaves as follows:
when a bell rings at $j$, if there is a $\eta$ particle
at $j$, then $\gamma(j)$ flips to $1-\gamma(j)$.  The process
$(\eta_t,\gamma_t)$ is Markovian and has $\nu^\rho\times\nu^{1/2}$ as
invariant measure. Again, the time-reversed process defined by
$(\eta^*_t,\gamma^*_t)=(\eta_{-t-},\gamma_{-t-})$ has the same law as the
space-reflection of $(\eta_t,\gamma_t)$: the jumps of $\eta^*_t$ go to the
left and the law of the spin-flip $\gamma^*_t$ remains the same. On the other
hand, given a trajectory of $((\eta_t,\gamma_t),\, t\ge 0)$ one can identify
$\omega$ as the space-time points $(j,t)$ such that 
either an $\eta$ particle arrives
at site $j$ at time $t$ or $\gamma$ flips at $j$ at time $t$.  The points
governing the reverse process are the time reflection of
$\Delta_\rho(\omega)$.  Since the reverse process has the same law as the
space-reflected HAD+spin-flip process, the points governing it must be
Poisson.

For any $t$, the dual points $\{(j,s)\in \Delta_\rho(\omega): s<t\}$
are the points governing the evolution
of the reverse process on the time interval $(-t,\infty)$
starting at
the configuration $(\eta^*_{-t}, \gamma^*_{-t})$,
and are independent of this configuration.
But this is just the configuration $(\eta_t, \gamma_t)$
so the independence holds as desired.
\square

\subsection{Multi-line HAD process} 
\label{sec:multilineHAD}

We now define a \emph{multi-line} process
$\alpha_t=(\alpha^1_t,\dots,\alpha^n_t)$ 
taking values in $\cX^n$.
It is again governed by a Poisson process $\omega$ on $\R\times\Z$.

Let $\rho^1,\dots,\rho^n\in(0,1)$.
Let $\omega^n=\omega$, and, recursively for $k=n-1,\dots,1$,
let $\omega^k=\Delta_{\rho^{k+1}}(\omega^{k+1})$.
From Proposition \ref{dp}, each $\omega^k$ is a Poisson process
of rate 1 on $\R\times\Z$.

Now let the ``$k$th line'' of the process,
$(\alpha_t^k,t\in\RR)$,
be $H_{\rho^k}(\omega^k)$, the HAD trajectory with density $\rho^k$
governed by the points $\omega^k$,
as provided by Proposition \ref{uniqpropos}.
Thus each line of the process is a HAD trajectory 
governed by the dual points produced from the line below.

Note also that, directly from the definition,
$(\alpha^1_t,\dots,\alpha^{n-1}_t)$ is a multi-line process with densities
$\rho^1,\dots,\rho^{n-1}$ and governed by $\omega^{n-1}$.
\begin{propos}
  \label{w2}
The multi-line HAD process $(\alpha_t,t\in\RR)$ is stationary,
and the distribution of $\alpha_t$ for each $t$
is the product measure 
$\nu=\nu^{\rho^1}\times\dots\times\nu^{\rho^n}$.
\end{propos}

\paragraph{\bf Proof}
By construction, the process is stationary and
the marginal distribution of 
$\alpha_t^k$
is $\nu^{\rho^k}$ for any $k$ and $t$.
So we need to show that, for any fixed $t$,
the configurations $\alpha_t^1,\alpha_t^2,\dots,\alpha_t^n$
are independent.

Let $2\leq k\leq n$. 
By Proposition \ref{dp},
the configuration $\alpha^k_t$
is independent of the set of dual points $(x,s)$
in $\Delta_{\rho^k}(\omega^k)$ such that $s<t$.

But the process $(\alpha^{k-1}_s, s\leq t)$
can be constructed as a function of precisely
this set of dual points; and then, recursively,
also the processes $(\alpha^{j}_s, s\leq t)$ for
each $1\leq j\leq k$.

In particular, we can construct 
$\alpha^{k-1}_t, \alpha^{k-2}_t, 
\dots, \alpha^{1}_t$ from the given set of dual points.

Thus for all $i$, the configuration $\alpha^k_t$ is independent of
$\alpha^{k-1}_t, \alpha^{k-2}_t, 
\dots, \alpha^{1}_t$. Hence all the $\alpha^k_t$
are independent as desired.
\square

\noindent
\textit{Remark:}
The dynamics of the multi-line process governed by $\omega$
can be explained in a more constructive (or ``local'') way.
Each bell $(j,s)\in\omega=\omega^n$ causes an $\alpha^n$ particle to jump to 
$j$, from $j^n$ say. This creates a bell
$(j^{n},s)$ in $\omega^{n-1}$,
which summons an $\alpha^{n-1}$ particle to $j^{n}$ from $j^{n-1}$,
causing a bell $(j^{n-1},s)$ in $\omega^{n-2}$ and so on.

\begin{figure}[htb]
\begin{center}
\input{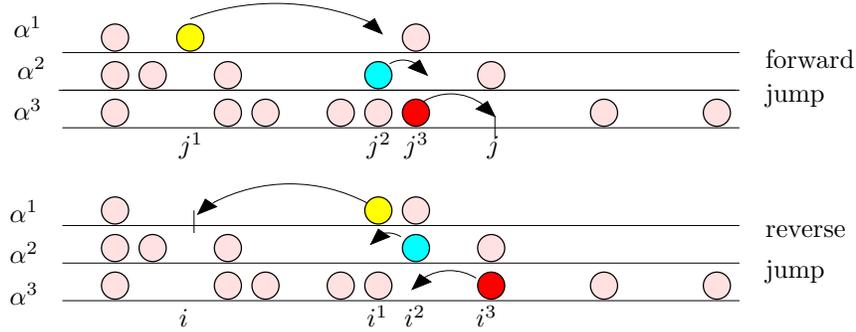}  
\caption{Local construction of the multi-line HAD process
and its time-reversal.} 
\label{rmp1}
\end{center}
\end{figure}

The time-reversal of this process, with respect to the equilibrium measure
$\nu$, can be described in the same way but with left and right
exchanged and also top and bottom exchanged.
When a Poisson bell rings at site $i$, the closest
$\alpha^1$ particle 
to the right of $i$ (including $i$) located at a site called
$i^1$ jumps to $i$. Then a bell 
rings at site $i^1$ for $\alpha^2$, and so on.
See Figure \ref{rmp1}. 
An alternative proof to Proposition \ref{w2} is to show 
directly 
that the process so
defined is the reverse process with respect to the product measure. 
We followed such a strategy 
for the case of the TASEP in \cite{FM-tasep}.

Note also that in the definition of the multi-line process,
and in Proposition \ref{w2}, we don't require
the densities $\rho^k$ to be increasing.

Now we wish to show that the image of 
the multi-line process under the map $T$ 
is the coupled process. 

\begin{propos}
\label{w1}
Let $0<\rho^1<\dots<\rho^n<1$, and
let $(\alpha_t,t\in\RR)$ be 
the multiline HAD trajectory governed by $\omega$
with densities $\rho^1,\dots,\rho^n$.
Let $\eta_t=T\alpha_t\in\Xnup$.
Then $(\eta^k_t, t\in\RR)$ 
is the HAD trajectory governed by $\omega$, with particle density $\rho^k$.
\end{propos}

\paragraph{\bf Sketch of Proof of Proposition \ref{w1}}
From the definition of $T$, we have
\begin{equation}
\label{Tcorres}
\eta_t^k=D^{(n-k+1)}\left(\alpha^k_t,\dots,\alpha^n_t\right).
\end{equation}
From Proposition \ref{Dnthings},
we know that $\eta_t^k$ has distribution $\nu^{\rho^k}$.
So we simply need to show that the RHS of \reff{Tcorres}
is a HAD trajectory governed by $\omega$.

Since $(\alpha^k,\dots,\alpha^n)$ is itself just a multi-line process
(with $n-k+1$ lines) governed by $\omega$, it is enough to 
show that, for any $n$, 
$D^{(n)}(\alpha^1_t,\dots,\alpha^n_t)$
is a HAD trajectory governed by $\omega$.

We argue by induction. 
From the definitions of $D^{(n)}$ and of the multi-line process,
the induction step is simple, using
\[
D^{(n)}(\alpha^1_t,\dots,\alpha^n_t)=
D^{(2)}\left(
D^{(n-1)}(\alpha^1_t,\dots,\alpha^{n-1}_t), \alpha^n_t
\right)
\]
and the fact that $(\alpha_t^1,\dots,\alpha_t^{n-1})$
is an $(n-1)$-line multiline process governed by $\omega^{n-1}$, as observed
just before Proposition \ref{w2}.

The base case $n=2$ remains. We use
the local description of the multi-line (in fact, two-line) process.
Let $(\alpha^1_t,\alpha^2_t)$ be the two-line process governed 
by $\omega$. Each mark $(x,s)$ in $\omega=\omega^2$ produces
a jump in the process $\alpha^2$ at time $s$,
and a corresponding dual point $(x',s)$ which becomes a mark 
in $\omega^1$. This mark in $\omega^1$ produces a jump in 
the process $\alpha^1$ at time $s$. One needs to verify that
the combination of the two jumps, in $\alpha^1$ and $\alpha^2$,
leads to a single HAD jump in the process $D(\alpha^1,\alpha^2)$,
equivalent to a mark at $x$. 
In the language used at \reff{a2}, we need to show
$D(A_{j'}\alpha^1,A_j\alpha^2)=A_j D(\alpha^1,\alpha^2)$.
This is not difficult to do
by checking a small number of cases. Arguing jump by jump in this way,
one obtains that $D(\alpha^1_t,\alpha^2_t)$ is a HAD
process governed by $\omega$ as required.
We give a proof along these
lines for the continuous-space case in \cite{FM-had}, and 
the same argument works here. 
\square

\paragraph{\bf The long range exclusion process}
\begin{figure}[htb]
\begin{center}
\input{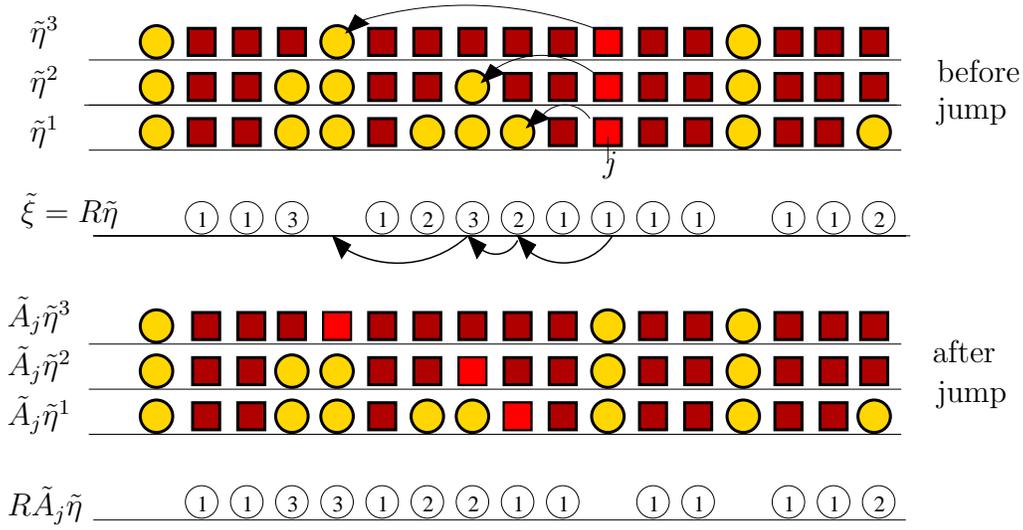}  
\caption{Jumps in coupled and multiclass 
LREP due to bell at $j$. LREP particles
  are represented by squares and empty sites by balls.}
\label{lrep}
\end{center}
\end{figure}
 The 
\textit{long range exclusion
process} (LREP) was introduced by Spitzer \cite{spitzer}. 
At rate one, a particle
located at site $x$ jumps to the first empty site found by a Markov chain with
transition jumps $p(.,.)$ starting at $x$. Consider empty sites of the HAD
process as particles and particles as empty sites. The resulting process
$\teta_t$ given by $\teta_t(x) = 1-\eta_t(x)$ is the 
LREP with transition matrix
$p(x,x-1) =1$ (Guiol \cite{Guiol}). The effect of a bell at $j$ is represented
by the map $\tA_j\teta(i) := 1-A_j\eta(i)$. In this simple case, the site is
just the first empty place to the left. 
The multiclass LREP $(\txi_t)$ has the
same distribution as the multiclass HAD process with classes reversed.  The
coupled LREP is defined by $\teta^k_t(x) = 1-\eta^{n+1-k}_t(x)$, where
$\eta_t=(\eta^1_t,\dots,\eta^n_t)$ is the coupled HAD.  The effect of the bell
at $j$ in the coupled LREP is given by the map
$\tC_j\teta=(\tA_j\teta^1,\dots,\tA_j\teta^n)$. The multiclass LREP is given by
$R\teta = \txi$.
In the multiclass LREP when a bell rings at $j$, the particle at $j$ jumps to
the closest particle to the left of it with higher class or empty;
simultaneously this particle jumps to the closest site to its left with higher
class or empty, and so on. 
The jumps finish when a particle jumps to an empty site.
See Figure \ref{lrep}.

\section{TASEP}
\label{stasep}

The totally asymmetric simple exclusion process, or TASEP, 
is a continuous-time Markov process in 
$\cX$ with the following dynamics. At
rate 1 if there is a particle at site $i$, it jumps one unit to the
left (if the site to the left is empty). We use here the same notations 
as in the case of the HAD process to describe the analogous 
quantities for the TASEP.
\begin{figure}[htb]
\begin{center}
\input{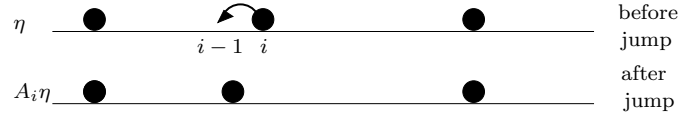}  
\caption{Jump in TASEP due to bell at $i$} 
\end{center}
\end{figure}
The generator of the process is given by
\begin{equation}
  \label{a40}
  Lf(\eta) = \sum_j [f(A_j\eta)-f(\eta)]
\end{equation}
where here (differently from the HAD definition \reff{a2}) 
\begin{equation}
  \label{a41}
  (A_j\eta)(k) = \left\{\begin{array}{ll} 
\eta(k) &\hbox{ if } k\notin\{ j-1,j\} \\ 
\max\{\eta(j-1), \eta(j)\} &\hbox{ if } k=j-1\\
\min\{\eta(j-1), \eta(j)\} &\hbox{ if } k=j.\\
\end{array}\right. 
\end{equation}
For the graphical construction of the process
we will use a system of Poisson points or marks,
this time on $\omega$ on $\R\times(\Z+\frac12 )$
(so that the bells now ring \emph{between} sites).
When a bell rings at site $x$, and there is a
particle at $x+\frac12 $ and a hole at $x-\frac12 $, 
the contents at sites $x+\frac12 $ and $x-\frac12 $ are ``interchanged''. 
\begin{figure}[htb]
\begin{center}
\input{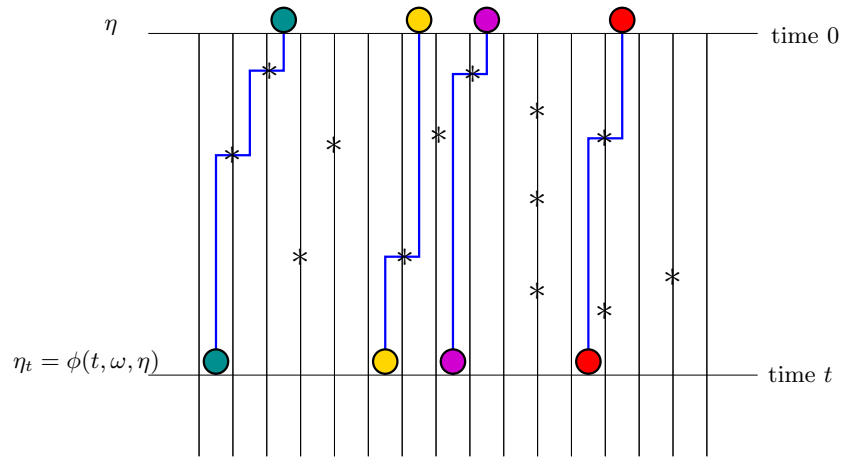}  
\caption{Graphical construction of TASEP. The * represent the events of the
  Poisson process $\omega$.}
\label{gct}
\end{center}
\end{figure}
The construction induces again a function $\phi$ of
$\omega$ and the initial configuration $\eta_0$:
\begin{equation}
  \label{aa4}
  (t,\omega,\eta)\mapsto\phi(t,\omega,\eta_0)
\end{equation}
and $\eta_t = \phi(t,\cdot,\eta_0)$ is the TASEP with initial configuration
$\eta_0$. We say that the points $\omega$ \emph{govern} the process
$\eta_t$. As at \reff{st} one has 
\begin{equation}
\label{st2}
\eta_t=\phi(t-s,\omega,\eta_s)
\end{equation}
for all $0\leq s<t<\infty$.

The Bernoulli product measures $\nu^\rho$ (and mixtures of them)
are again invariant for the TASEP. (In addition,
certain \emph{blocking measures} are also invariant;
these measures are concentrated on a single configuration
with only particles to the left of some site and only holes to its
right).
 
If a trajectory $(\eta_t,t\in\RR)$ satisfies \reff{st}
for all $-\infty<s<t<\infty$ then we again say 
that it is governed by $\omega$. 
The following result is analogous to Proposition \ref{uniqpropos}:
\begin{propos}
\label{TASEPuniqpropos}
Let $\rho\in(0,1)$.
Then there exists an essentially unique function $H_\rho$ 
mapping elements $\omega$ of $\Omega$ to trajectories
$(\eta_t,t\in\R)$ such that:
\begin{itemize}
\item[(i)]
The induced law of $(\eta_t,t\in\R)=H_\rho(\omega)$ is stationary in time.
\item[(ii)]
The marginal law of $\eta_t$ for each $t$ is space-ergodic with particle density
$\rho$.
\item[(iii)]
With probability 1, $(\eta_t,t\in\R)$ is a TASEP evolution governed by $\omega$.
\end{itemize}
Then in fact the marginal law of $\eta_t$ for each $t$ is $\nu^\rho$.
\end{propos}

\subsection{Coupled and multiclass TASEP}

\begin{figure}[htb]
\begin{center}
\input{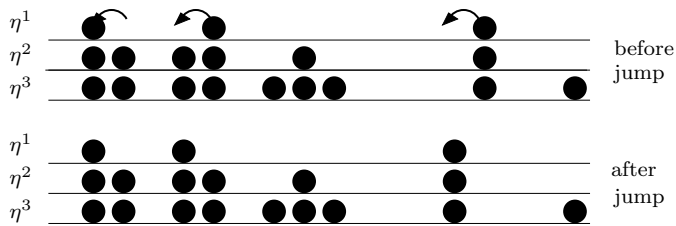}  
\caption{Coupling in TASEP} 
\label{tasep-coupling}
\end{center}
\end{figure}

The basic coupling between $n$ TASEPs with initial configurations
$\eta=\eta^1_0,\dots,\eta^n_0$ is given by
$\eta_t=(\eta^1_t,\dots,\eta^n_t)=\phi^{(n)}(t,\omega,\eta_0)$,
where
$\phi^{(n)}(t,\omega,\eta_0)^k=\phi(t,\omega,\eta^k_0)$.
See figure \ref{tasep-coupling}, where the effects of three possible  
Poisson bells are indicated.
If $\eta_0^1\leq\dots\leq\eta_0^n$,
then this ordering is preserved by the coupling. 
Thus we obtain a coupled process $(\eta_t)$, governed by $\omega$,
taking values in $\Xnup$.

\begin{figure}[htb]
\begin{center}
\input{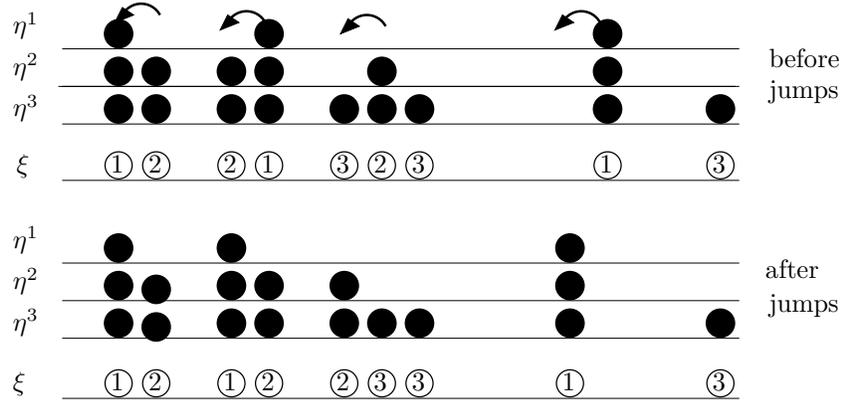}  
\caption{Coupled and multiclass TASEP}
\label{mct} 
\end{center}
\end{figure}

The equivalent multiclass process 
$(\xi_t)$ with values in $\cY_n$ is then obtained by putting 
$\xi_t=R\eta_t$.
The evolution of this multiclass TASEP
is more intuitive than in the case of the HAD process. 
At the ring of the bell at $x$, particles of
lower class at site $x+\frac12 $ jump over particles of higher class at $x-\frac12 $,
exchanging places. See four examples of jumps in Figure \ref{mct}.
\begin{thm}
  \label{t2}
Under the conditions of Theorem \ref{t1}, $\pi$, the distribution of
$\eta=T\alpha$ is
invariant for the coupled TASEP $(\eta_t)$,
and $\mu$, the law of $M\alpha$, is invariant
for the multiclass TASEP $(\xi_t)$.
\end{thm}
The strategy to prove this theorem is the same as for Theorem \ref{t1}. The
differences come in the definitions
of dual points and 
of the multi-line process.
The key steps are given 
by Proposition \ref{ww2} 
and Proposition \ref{ww1},
which play the roles played by Propositions \ref{w2} and \ref{w1} 
in the case of the HAD process.

\subsection{Dual points in TASEP}

\begin{figure}[htb]
\begin{center}
\input{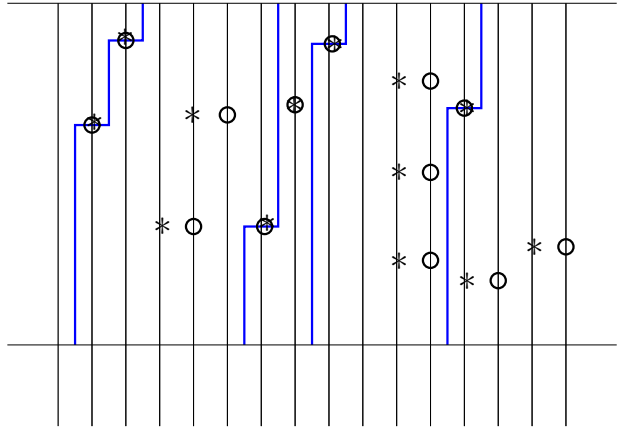}  
\caption{The circles represent the dual points of the TASEP trajectory of
  Figure \ref{gct}.
As in Figure \ref{ml1b},
turning the picture upside-down exchanges the roles of 
circles and stars.}
\label{dpt} 
\end{center}
\end{figure}

We now define the \emph{dual points} for the case of TASEP.
Let the density $\rho$ be fixed and let $\omega$ 
be a Poisson process on $\R\times(\Z+\frac12 )$.
Let $(\eta_t,t\in\RR)$ be the TASEP trajectory $H_\rho(\omega)$ governed 
by $\omega$, as provided by Proposition \ref{TASEPuniqpropos}.
Now define the dual points $\Delta_\rho(\omega)$ by
\begin{equation}
\label{c10}
\Delta_\rho(\omega)=
\{
(x,t)\in\omega:\,\eta_{t-}(x+\tfrac12 )=1\}
\cup
\{(x+1,t):\,(x,t)\in\omega \hbox{ and } \eta_{t-}(x+\tfrac12 )=0\}. 
\end{equation}
See Figure \ref{dpt}.
\begin{propos}
  \label{c11}
Let $\omega$ be a Poisson process in 
$\R\times\Z$, and let $\Delta_\rho(\omega)$ be the dual 
points for the TASEP trajectory with particle density $\rho$
governed by $\omega$.
Then $\Delta_\rho(\omega)$ is also a Poisson process in $\R\times\Z$.
Furthermore
$\{(x,s)\in \Delta_\rho(\omega):\, s<t\}$, 
the set of dual points earlier than 
$t$, is independent of the configuration $\eta_t$.
\end{propos}

\paragraph{\bf Proof} This is a kind of Burke's theorem like Proposition
\ref{dp}. Again the TASEP trajectory 
does not determine the points of $\omega$,
so we introduce spin-flip processes to
mark the missed points. There are two types of 
points missed by the trajectory:
those space-time points $(x,t)$ 
such that $\eta_{t-}(x+\frac12 )=0$ and those space-time points $(x,t)$
such that $\eta_{t-}(x+\frac12 )=\eta_{t-}(x-\frac12)=1$.
We mark these two types with different spin-flip process. 
Let $\gamma_t$ be a process taking values in $\cX=\{0,1\}^\Z$
which behaves as follows: 
when there is an $\omega$ point at $(x,t)$, if there is no $\eta_{t-}$
particle at $x+\frac12 $, then the value of $\gamma$ at $(x+\frac12 ,t)$ flips
so that 
$\gamma_{t}(x+\frac12 )=1-\gamma_{t-}(x+\frac12 )$.  
Let $\zeta_t$ be a spin-flip process taking values in
$\{0,1\}^{\Z+\frac12}$ with the following behavior: 
when there is an $\omega$ point at $(x,t)$, if both 
$x-\frac12 $ and $x+\frac12 $ are occupied by $\eta_{t-}$
particles, then the value of $\zeta$ at $(x,t)$ flips so that
$\zeta_{t}(x)=1-\zeta_{t-}(x)$. 
Given the evolution
$( (\eta_t,\gamma_t,\zeta_t),\, t\in \R)$,
the points 
$\omega$ can be recovered by
\begin{equation}
  \label{c15}
  \omega = \{(x,t):\,  (\eta_{t-}(x+\tfrac12 ),\gamma_{t-}(x+\tfrac12 ),
\zeta_{t-}(x)) 
  \neq (\eta_{t}(x+\tfrac12 )\eta_{t-}(x+\tfrac12 ),\gamma_{t}(x+\tfrac12 ),
\zeta_{t}(x))\}
\end{equation}
We now proceed as in the proof of Proposition \ref{dp}.
The process $( (\eta_t,\gamma_t,\zeta_t),\, t\in \R)$ 
is stationary with time-marginal product measure
$\nu^\rho\times\nu^{1/2}\times\nu^{1/2}$. The reverse process with respect to
this measure is defined by $(\eta^*_t,\gamma^*_t,\zeta^*_t) =
(\eta_{-t-},\gamma_{-t-},\zeta_{-t-})$. 
Then the law of this time-reversal is the same
as the law of the space-reversal.
(In particular, the first coordinate of the reverse
process performs a 
TASEP with jumps to the \emph{right}, 
while the other two coordinates
have the same spin flip distribution as the forward process).
The points governing the reverse process are the time-reflection
of $\Delta_\rho(\omega)$, which is therefore also a Poisson process
on $\R\times(\Z+\frac12 )$.
Independence is shown as in Proposition \ref{dp}. \square

\subsection{Multi-line TASEP}

We now define a multi-line TASEP 
$\alpha_t=(\alpha_t^1,\dots,\alpha_t^n)$ taking values in $\cX^n$
and governed by Poisson points $\omega$ on $\R\times(\Z+\frac12)$.
The definition is analogous 
to that of the multi-line HAD in Section \ref{sec:multilineHAD}.

Let $\rho^1,\dots,\rho^n\in(0,1)$.
Let $\omega^n=\omega$, and, recursively for $k=n-1,\dots,1$,
let $\omega^k=\Delta_{\rho^{k+1}}(\omega^{k+1})$.
Now let the $k$th line of the process,
$(\alpha_t^k,t\in\RR)$,
be $H_{\rho^k}(\omega^k)$, 
the TASEP trajectory with density $\rho^k$
governed by the points $\omega^k$,
as provided by Proposition \ref{TASEPuniqpropos}.

Using Proposition \ref{c11},
an argument analogous to the proof of Proposition 
\ref{w2} now gives the following result:
\begin{propos}
\label{ww2}
The multi-line TASEP $(\alpha_t,t\in\RR)$ is stationary,
and the distribution of $\alpha_t$ for each $t$ is 
the product measure $\nu=\nu^{\rho^1},\dots,\nu^{\rho^n}$.
\end{propos}

The proof of Theorem \ref{t2} is completed by the following 
result, analogous to 
Proposition \ref{w1}:
\begin{propos}
\label{ww1}
Let $0<\rho^1<\dots<\rho^n<1$, and
let $(\alpha_t,t\in\RR)$ be 
the multiline TASEP trajectory governed by $\omega$
with densities $\rho^1,\dots,\rho^n$.
Let $\eta_t=T\alpha_t\in\Xnup$.
Then $(\eta^k_t, t\in\RR)$ 
is the TASEP trajectory governed by $\omega$, with particle density $\rho^k$.
\end{propos}

\paragraph{\bf About the proof of Proposition \ref{ww1}}
The induction argument is the same as for Proposition \ref{w1}.
The case-by-case checking for $n=2$ must now be done 
for the TASEP dynamics. 
We have done this in \cite{FM-tasep}. 
\square

\section{Other dynamics}
\label{sdynamics}
There are other examples of dynamics on $\cX$ for which the associated
multiclass processes also have the family of measures $\mu$ 
as invariant distributions.

For example, consider the \emph{sequential TASEP}.
This is a discrete-time Markov chain with values in $\ZZ$.
At each time step, each 
particle tries to jump left 
with probability $p$,
succeeding if the site to its left is empty.
Updates are carried out sequentially from left to right
(so for example a particle may jump into a space which 
is only vacated at the same time-step).
Now the governing points $\omega$ have Bernoulli product
measure on $\Z\times\Z$ (and the same is true
of the dual points, appropriately defined).
An analogous method of proof via a multi-line process
shows that $\mu$ is invariant for the multiclass process.

The same is true for a form of sequential TASEP with updates from
right to left. Now a particle may not jump immediately
into a vacated space, but the same particle may jump
several times at the same time-step (because of several
neighbouring points $(x,t)$, $(x-1,t), \dots$ in 
the governing configuration $\omega$).
In fact this process is dual to the one in the previous
paragraph, under exchange of hole and particle and of left and right.
The measure $\mu$ is invariant for the multiclass version,
and similarly in the case of various discrete-time versions
of the HAD process.

However, consider instead the \emph{parallel TASEP}, again 
in discrete time. All sites are updated simultaneously; 
now jumps are only allowed at sites containing a particle
with a hole to its left before any other update occurred.
In particular, jumps at two neighbouring sites cannot occur 
at the same time-step.
In this case, the basic coupling does not 
even preserve ordering of configurations,
so that the multiclass process cannot be defined in the same way.
Note also that product measure $\nu^\rho$ is no longer
invariant for the parallel TASEP.

Consider also the asymmetric simple exclusion process (ASEP)
in continuous time,
in which each particle tries to jump left at rate $p$ 
and right at rate $1-p$.
Product measure $\nu^\rho$ is invariant for the process; 
however, unless $p=1$, the measure $\mu$ is no longer invariant for the 
multiclass process. A very interesting question
is whether the invariant multiclass measures
of these more general ASEPs could be constructed
using an approach related to the one described here
for the TASEP.

\section{Multiclass Burke's Theorem}
\label{sburke}

For each $n$ and densities $0<\rho^1<\dots<\rho^n<1$,
we have constructed a measure $\mu=\mu^{(n)}_{\rho^1,\dots,\rho^n}$
on $\cY_n=\{1,2,\dots,n+1)^\Z$
which is invariant for the multiclass HAD process and the 
multiclass TASEP.
A configuration from $\mu^{(n)}_{\rho^1,\dots,\rho^n}$
has density $\rho^1$ of first-class particles
and density $\rho^k-\rho^{k-1}$ of $k$th class particles,
for $k=2,\dots,n$.

Fix $n$ and $\rho^1,\dots,\rho^n$, and let $m<n$.
Let $\xi^{(n)}$ be distributed according to 
$\mu^{(n)}_{\rho^1,\dots,\rho^n}$,
and $\xi^{(m)}$ according to 
$\mu^{(m)}_{\rho^1,\dots,\rho^m}$.
Then a nice property of this family of distributions is that
\begin{equation}
  \label{b1}
\hbox{$\xi^{(m)}$ has the same distribution as $[\xi^{(n)}]^m$},
\end{equation}
where $[\xi^{(n)}]^m$ is
the truncated configuration defined by
\begin{equation*}
[\xi^{(n)}]^m(i) = 
\min\left\{
\xi^{(n)}(i),m+1
\right\}.
\end{equation*}
Putting $n=m+1$, we obtain the following statement
in the context of the tandem queueing system described in Section \ref{sim}:
the $m$-class input process to queue $m$ has
the same law as the $m$-class departure process
from the same queue. Thus the measure
$\mu^{(m)}_{\rho^1,\dots,\rho^m}$
is a \textit{fixed point} for a 
discrete-time $./M/1$ priority queue with $m$ classes
(here $./M/1$ denotes a queue whose 
sequence of potential service times is a Bernoulli process).

This may be called a \textit{multiclass Burke's theorem}.
The original form of Burke's theorem,
in this discrete-time setting,
states that a Bernoulli process is a fixed point for a 
(one-class) $./M/1$ queue; this 
is the statement \reff{b1} specialized to the case $n=2$ and $m=1$
(see e.g. \cite{Reich}, and \cite{HsuBurke} for the discrete-time case).
Property \reff{b1} is easy to deduce from results above, 
using the uniqueness of the invariant measure for the
coupled process 
with given particle densities
(see \cite{FM-had} for the equivalent argument in continuous time).
A more direct proof can be found in \cite{MarPra}, using properties
of invariance of the law of the departure process from a 
tandem queueing system under interchange of the order of the queues.

\section*{Acknowledgements}

Thanks to Eric Cator, Sheldon Goldstein and Herbert Spohn for enjoyable
discussions.

This paper is partially supported by the Brazil-France Agreement, FAPESP, CNPq
and PRONEX. PAF thanks hospitality and support from IHES and Laboratoire de
Probabilit\'es of Universit\'e de Paris 7. JBM thanks IME-USP and the FAPESP.


\bigskip

\parindent0pt
\parbox{0.54\linewidth}
{
\textsc
{Instituto de Matem\'atica e Estat\'istica,\\
Universidade de S\~ao Paulo,\\
Caixa Postal 66281,\\
05311-970 S\~ao Paulo,\\
Brazil}\\
\texttt{pablo@ime.usp.br}\\
\texttt{http://www.ime.usp.br/$\tilde{\,\,\,}$pablo}}
\hfill
\parbox{0.48\linewidth}
{\textsc
{LIAFA,\\
CNRS et Universit\'e Paris 7,\\
2 place Jussieu (case 7014),\\
75251 Paris Cedex 05\\
France}\\
\texttt{James.Martin@liafa.jussieu.fr}\\
\texttt{http://www.liafa.jussieu.fr/$\tilde{\,\,\,}$martin}}

\end{document}